%% file: main.tex
\documentclass[sigconf,nonacm]{acmart}

\AtBeginDocument{%
  \providecommand\BibTeX{{%
    \normalfont B\kern-0.5em{\scshape i\kern-0.25em b}\kern-0.8em\TeX}}}

\usepackage[breakable]{tcolorbox}

\usepackage{multirow}
\usepackage{subcaption}
\usepackage{caption}
\usepackage{tabularx}
\usepackage{pifont}
\usepackage{xcolor}
\usepackage{makecell}

\begin{document}

\title[ResearchCube: Multi-Dimensional Trade-off Exploration for Research Ideation]{ResearchCube: \\ Multi-Dimensional Trade-off Exploration for Research Ideation}

\author{Zijian Ding}
\affiliation{%
  \institution{University of Maryland, College Park}
  \country{}}\email{ding@umd.edu}

\author{Fenghai Li}
\affiliation{%
  \institution{University of Illinois Urbana-Champaign}
  \country{}}\email{max7@illinois.edu}

\author{Ziyi Wang}
\affiliation{%
  \institution{University of Maryland, College Park}
  \country{}}\email{zoewang@umd.edu}

\author{Joel Chan}
\affiliation{%
  \institution{University of Maryland, College Park}
  \country{}}\email{joelchan@umd.edu}

\begin{abstract}
Research ideation requires navigating trade-offs across multiple evaluative dimensions, yet most AI-assisted ideation tools leave this multi-dimensional reasoning unsupported, or reducing evaluation to unipolar scales where "more is better". We present ResearchCube, a system that reframes evaluation dimensions as bipolar trade-off spectra (e.g., theory-driven vs. data-driven) and renders research ideas as manipulable points in a user-constructed 3D evaluation space. Given a research intent, the system proposes candidate bipolar dimension pairs; users select up to three to define the axes of a personalized evaluation cube. Four spatial interactions--AI-scaffolded dimension generation, 3D navigation with face snapping, drag-based idea steering, and drag-based synthesis--enable researchers to explore and refine ideas through direct manipulation rather than text prompts. A qualitative study with 11 researchers revealed that (1) bipolar dimensions served as cognitive scaffolds that externalized evaluative thinking and offloaded working memory, (2) the spatial representation provided a sense of agency absent in chatbot-based AI tools, (3) participants desired fluid transitions across dimensionality levels---from single-dimension focus to more than three dimensions, and (4) a productive tension emerged between AI-suggested starting dimensions and users' evolving desire for control. We distill these findings into design implications for multi-dimensional research ideation tools, including progressive dimensional control, fluid dimensionality, and transparent synthesis with provenance.
\end{abstract}


\begin{teaserfigure}
  \centering
  \includegraphics[width=0.85\textwidth]{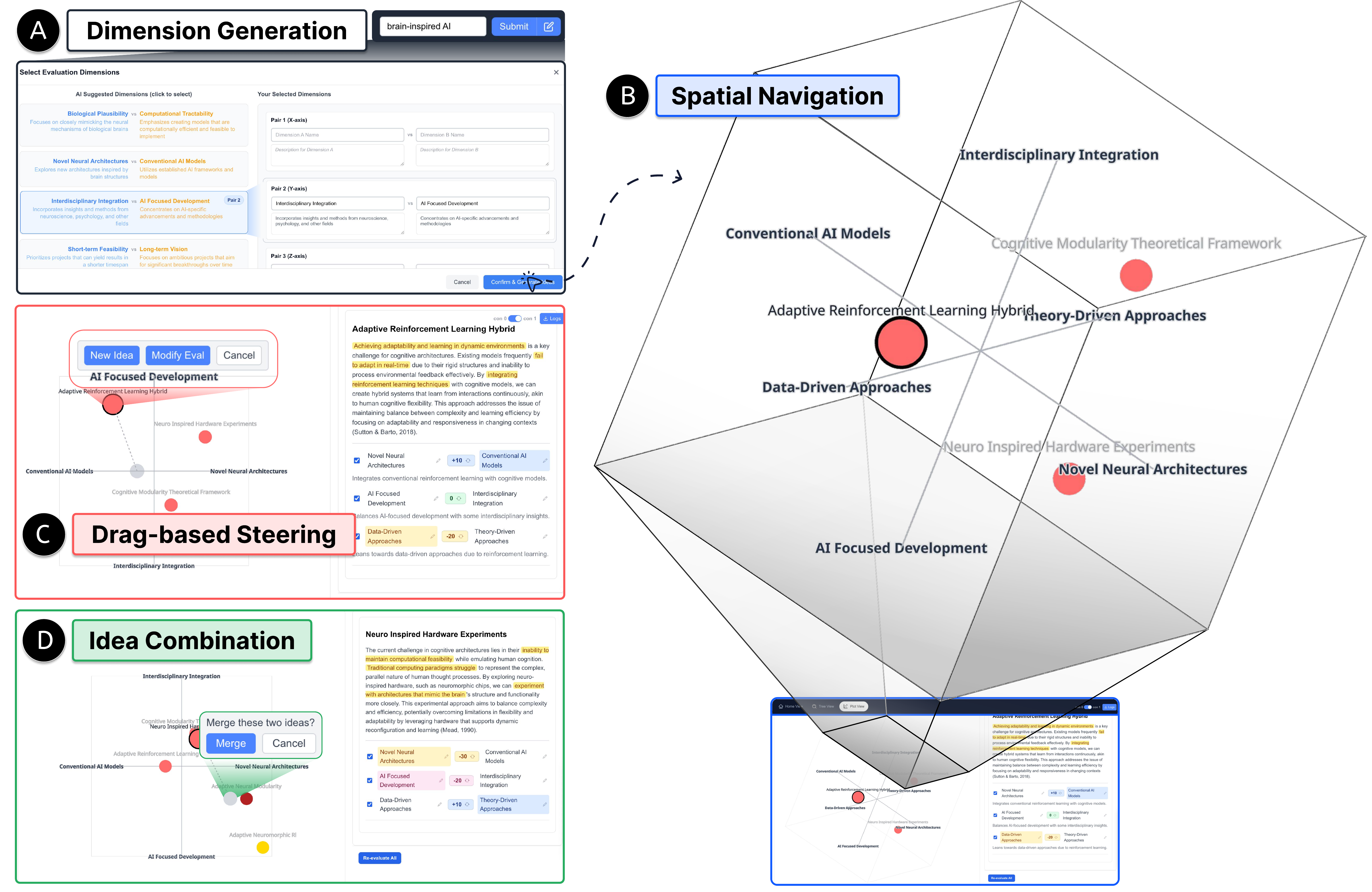}
  \caption{ResearchCube renders research ideas as interactive nodes in a 3D evaluation space. Each axis represents a user-selected bipolar trade-off dimension (e.g., theory-driven vs. data-driven). Users explore ideas by rotating the cube, and refine them by dragging nodes toward desired trade-off balances.}
  \Description{3D evaluation space showing research ideas as spheres positioned along three bipolar trade-off axes, with axis labels at each endpoint.}
  \label{fig:teaser}
\end{teaserfigure}

\maketitle

\input{sections/1-introduction}
\input{sections/2-relatedWork}
\input{sections/3-systemDesign}
\input{sections/4-studyDesign}
\input{sections/5-results}
\input{sections/6-discussion_conclusion}

\bibliographystyle{ACM-Reference-Format}
\bibliography{reference}

\input{sections/7-appendix}

\end{document}

%% file: sections/1-introduction.tex
\section{Introduction}
\label{sec:intro}

Research ideation involves navigating a vast space of possibilities along multiple evaluative dimensions---novelty, feasibility, impact, methodological rigor, and domain fit, among others. Navigating this multi-dimensional space is part of what makes for creative ideas, particularly for ``wicked,'' open-ended problems with no obvious global optimum but rather negotiations and trade-offs. While we have good systems for exploring ideas based on facets~\cite{puIdeaSynthIterativeResearch2024, radenskyScideatorHumanLLMScientific2025} or spatial manipulation~\cite{suhLuminateStructuredGeneration2024, chungPatchviewLLMPoweredWorldbuilding2024}, evaluative multi-dimensionality is seldom an explicit focus---dimensionality is often left implicit in favor of foregrounding idea facets or linear chats. This lack of explicit support leaves important and effortful work to be done primarily using the researcher's internal cognitive resources, missing opportunities for deeper human-AI collaboration on research ideation.

For example, prior work on Direct Intent Manipulation (DIM)~\cite{dingDirectIntentManipulation2025} demonstrated that drag-based steering in a 2D canvas helps researchers iterate on ideas more fluidly than prompt-only interfaces. However, DIM and similar systems evaluate ideas on fixed unipolar dimensions (e.g., novelty scored 0--100), implicitly treating each criterion as ``more is better.'' Research evaluation is rarely so straightforward: for example, high novelty might come at the cost of feasibility; broad scope might sacrifice depth. These interdependencies are hard to be captured by independent unipolar scales where ``more is better". A more nuanced approach would reframe each axis as a trade-off spectrum with meaningful poles at both ends: for example, consider
a researcher exploring ``AI for mental health": relevant trade-offs
might include individual-tailored vs. population-level interventions,
privacy-preserving vs. data-rich approaches, or clinically validated
vs. exploratory methods; none of which reduce to simple ``more is
better" scales.

Spatial representations offer a natural medium for externalizing multi-dimensional evaluation: position encodes trade-off balance and axes carry explicit semantic
meaning \cite{leeEvaluatingHumanLanguageModel2024}. A semantically
grounded coordinate system, where each axis represents an explicit evaluative trade-of, could make the evaluative structure of the
ideation space directly visible and manipulable.

This paper introduces \textbf{ResearchCube}\footnote{Open-source implementation available at: \url{https://github.com/[anonymized]} (link will be provided upon acceptance).}, a system for multi-dimensional trade-off exploration that renders research ideas as directly manipulable nodes in a spatial canvas parametrized by user-selected bipolar dimensions. This bipolar framing makes trade-offs explicit and supports more nuanced positioning of ideas. To help users construct their evaluation space, ResearchCube provides AI-assisted trade-off dimension generation: given a research intent, the system proposes candidate bipolar dimension pairs, from which users select up to three to define the axes of an evaluation space (up to 3D). This scaffolding addresses the cold-start problem---users who lack clear evaluative frameworks can bootstrap from AI suggestions while retaining full control over which trade-offs matter for their context.

We use ResearchCube as a design probe to explore the following research questions:
\begin{itemize}
    \item \textbf{RQ1.} How should a research ideation space be designed so that researchers can externalize, compare, and navigate evaluative trade-offs among candidate ideas?
    \item \textbf{RQ2.} What interaction mechanisms enable researchers to fluidly iterate on and synthesize ideas within such a space?
\end{itemize}

Using a combination of design iterations and an empirical design probe study with 11 researchers, we make the following contributions that address our research questions:
\begin{enumerate}
    \item \textbf{The design concept of bipolar trade-off dimensions as cognitive scaffolds.} We propose representing each evaluative criterion as a bipolar spectrum (e.g., theory-driven $\leftrightarrow$ data-driven) rather than a unipolar scale (e.g., novelty 0--100). Through our empirical study, we find that these bipolar dimensions function as cognitive scaffolds that externalize evaluative thinking and offload working memory, and that rendering them spatially restores a sense of agency absent in chatbot-based AI tools.
    \item \textbf{The design concept of drag-based steering and spatial synthesis for idea refinement.} We design two spatial interactions, drag-to-steer and drag-to-synthesize, that let researchers express revision intent through direct manipulation rather than text prompts. Our study reveals that drag-based steering enables spatial exploration of under-explored regions, though it demands cognitive investment, and that idea synthesis faces challenges around content transparency and convergence.
    \item \textbf{ResearchCube}, a proof-of-concept system that instantiates these design concepts in a 3D evaluation cube coupled with a provenance-preserving tree view. The system supports AI-scaffolded dimension generation, progressive idea streaming, proximity-based synthesis, and fragment extraction for non-linear recombination.
    \item \textbf{Empirical insights} ($n=11$) into how researchers engage with multi-dimensional trade-off spaces for ideation, including the finding that participants desire fluid transitions across dimensionality levels and that a productive tension emerges between AI-suggested starting dimensions and users' evolving desire for control.
\end{enumerate}

%% file: sections/2-relatedWork.tex
\section{Related Work}


\subsection{AI-Powered Research Ideation}

Recent studies have demonstrated that LLMs can generate and evaluate research ideas at levels approaching human expert agreement. Si et al.~\cite{siCanLLMsGenerate2024} found that evaluations using Claude-3.5 achieved a 53.3\% pairwise agreement rate compared to 56.1\% among NLP researchers. Baek et al.~\cite{baekResearchAgentIterativeResearch2025} showed that human-model agreement was higher for pairwise comparisons (0.71) than for scoring-based evaluations (0.64 vs.\ human-human 0.83), suggesting that the evaluation method itself matters. These systems typically evaluate ideas across multiple dimensions, including novelty, feasibility, impact, and clarity~\cite{siCanLLMsGenerate2024, yangLargeLanguageModels2024, baekResearchAgentIterativeResearch2025}, revealing that research ideation inherently involves navigating a high-dimensional evaluative space. Recent systems support iterative idea development through various mechanisms: IdeaSynth~\cite{puIdeaSynthIterativeResearch2024} enables facet-based idea combination, allowing researchers to merge concepts from different sources, and Scideator~\cite{radenskyScideatorHumanLLMScientific2025} supports scientific ideation through structured recombination of research components. We extend these AI-supported research ideation systems by making the evaluative dimensions along which ideas differ remain implicit in these systems explicit and manipulable as evaluative tradeoffs in a shared spatial representation.

\subsection{Constructing Multi-Dimensional Research Spaces}

A challenge in human exploration and evaluation of research ideas is the high-dimensional nature of the research space, where novelty, feasibility, impact, and other criteria often interact and compete. Presenting this high-dimensional space to researchers in a way that supports effective reasoning about trade-offs is even more challenging, and remains relatively underexplored. Prior work on textual interfaces for ideation has shown that listing example solutions can constrain creative exploration by anchoring users to existing ideas~\cite{chanFormulatingFixatingEffects2024}, whereas spatially contextualized examples positioned within a solution space can foster more diverse and exploratory thinking~\cite{chanFormulatingFixatingEffects2024}. This suggests that spatializing the evaluative structure of research ideas, rather than presenting them as flat text lists, could better support creative ideation. Among existing systems for spatial ideation, Luminate~\cite{suhLuminateStructuredGeneration2024} addresses the cold-start problem through AI-assisted dimension generation, proposing attribute dimensions based on user-provided examples for design exploration. Yet Luminate's dimensions emphasize attribute combinations rather than explicit trade-offs. Sensecape~\cite{suhSensecapeEnablingMultilevel2023} enables multi-level abstraction of LLM-generated content, but does not provide evaluative dimensions that users can directly manipulate to steer generation. Neither system frames dimensions as bipolar spectra where meaningful trade-offs can be spatially navigated. Once a multi-dimensional research space has been established, a complementary challenge arises: designing interaction mechanisms that allow researchers to fluidly explore, steer, and recombine ideas within that space. ResearchCube is trying to address both challenges: it supports constructing a customized multi-dimensional evaluation space through AI-scaffolded dimension generation, and provides spatial interaction mechanisms for fluidly exploring and refining ideas within that space.

\subsection{Direct Manipulation for Human-AI Ideation}

Preserving human agency is an ongoing challenge in human-AI collaboration such as AI-assisted ideation~\cite{heerAgencyAutomationDesigning2019}. When AI systems generate outputs autonomously, users may feel like passive recipients rather than active collaborators. Shared representations that both human and AI can inspect and modify offer one path toward agency-preserving interaction~\cite{heerAgencyAutomationDesigning2019}. Direct manipulation, continuous representation of objects, physical actions instead of syntax, and rapid reversible operations~\cite{shneidermanFutureInteractiveSystems1982}, has proven effective for complex cognitive tasks~\cite{hutchinsDirectManipulationInterfaces1985}. Spatial reasoning provides an intuitive substrate for expressing nuanced distinctions~\cite{gardenforsConceptualSpaces2000}, as humans naturally think in terms of proximity, direction, and relative position. Semantic interaction systems like ForceSPIRE~\cite{endertSemanticInteractionVisual2012} leverage this capacity by mapping drag gestures to updates in underlying analytical models, allowing users to steer document clustering through spatial arrangement. Recent work applies these principles to generative AI: PatchView~\cite{chungPatchviewLLMPoweredWorldbuilding2024} enables drag-based manipulation along opposing dimensions for worldbuilding, Direct Intent Manipulation (DIM)~\cite{dingDirectIntentManipulation2025} showed that dragging ideas toward target positions can steer research ideation more fluidly than prompt-only interfaces, and Toyteller~\cite{chungToytellerAIpoweredVisual2025} turns story fragments into interactive objects that foster expressive ideation through spatial play. ResearchCube builds on these foundations by combining direct manipulation with bipolar trade-off dimensions tailored for research ideation. Unlike prior systems that treat dimensions as independent attributes, ResearchCube frames each axis as a bipolar trade-off with meaningful poles at both ends, making tensions visible and manipulable. Evaluative dimensions serve as both organizational scaffolds and control surfaces: the dimensional structure that organizes ideas also serves as the interface for modifying them. 


%% file: sections/3-systemDesign.tex
\section{System Design}
\label{sec:system}

\begin{figure*}[t]
  \centering
  \includegraphics[width=\textwidth]{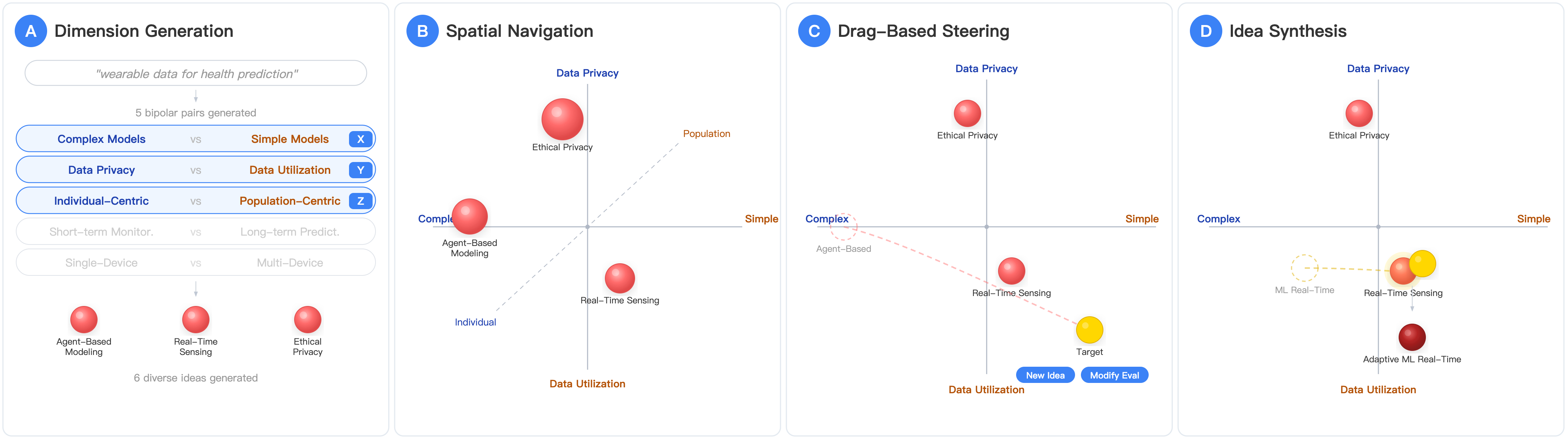}
  \caption{Usage senario of ResearchCube's four primary interactions, using data from P02 exploring ``wearable data for health prediction.'' (A)~Dimension generation: the system proposes five bipolar trade-off pairs; the user selects three (Complex Models vs.\ Simple Models, Data Privacy vs.\ Data Utilization, Individual-Centric vs.\ Population-Centric) to define a 3D evaluation space. (B)~Spatial navigation: six generated ideas are rendered as nodes in the 3D space; node size encodes depth along the Z-axis. (C)~Drag-based steering: the user drags Agent-Based Modeling toward the Simple + Data Utilization quadrant to express a revision intent; the system generates a new idea at the target position. (D)~Idea synthesis: the newly generated ML Real-Time Processing is dragged onto Real-Time Sensing to merge them, producing Adaptive ML Real-Time Health.}
  \Description{Four-panel usage scenario showing dimension generation, 3D spatial navigation, drag-based steering, and idea synthesis using participant data.}
  \label{fig:workflow}
\end{figure*}

ResearchCube supports four primary interactions organized around two complementary goals. The first two—dimension generation and spatial navigation—help users construct and orient themselves within a personalized multi-dimensional evaluation space. The second two—drag-based steering and idea synthesis—enable users to fluidly explore, refine, and recombine ideas within that space. The system also supports fragment-based recombination alongside full-idea synthesis given that researchers might identify promising elements distributed across multiple candidate ideas.

\subsection{Interaction Design}
\label{sec:interactions}


\subsubsection{Dimension Generation (Figure~\ref{fig:workflow}A)}
A challenge in multi-dimensional ideation is that users often lack a clear evaluative framework before they begin exploring. While Luminate~\cite{suhLuminateStructuredGeneration2024} introduced AI-assisted dimension generation for design exploration, ResearchCube extends this approach to research ideation with explicitly bipolar trade-off dimensions. To address this cold-start problem, the system generates candidate trade-off dimension pairs based on the user's initial research intent. When a user enters an intent (e.g., ``reinforcement learning for user intent prediction''), the system proposes five bipolar dimension pairs, each representing a meaningful trade-off spectrum relevant to the research domain. For example:
\begin{itemize}
    \item Model Complexity: simple vs. complex
    \item Methodology: theory-driven vs. data-driven
    \item Scope: narrow domain vs. broad generalization
    \item Novelty: incremental vs. paradigm shift
    \item Validation: simulation vs. real-world deployment
\end{itemize}
The interface presents these suggestions in a two-column selection panel: the left column displays AI-generated dimension pairs with descriptions explaining each pole; the right column shows three axis slots (X, Y, Z) where users assign their chosen pairs. Users may select one, two, or three pairs depending on the desired complexity of exploration. Once dimensions are confirmed, the system generates diverse seed ideas: the system prompt is tuned to generate ideas that vary broadly across the selected dimensions of the evaluation space, to help ensure broad initial coverage. Each idea is scored and positioned according to its trade-off balance, with results streaming progressively into the space.

\subsubsection{Spatial Navigation (Figure~\ref{fig:workflow}B)}
Navigating a 3D space on a 2D screen presents interaction challenges: free rotation can disorient users, and dragging in arbitrary orientations creates ambiguity about which dimensions are being modified. The system addresses this through \emph{camera snapping}. Users can freely rotate the evaluation cube to inspect ideas from any angle. However, when rotation stops, the camera automatically snaps to the nearest orthogonal face---front, back, left, right, top, or bottom---aligning the view with exactly two axes while locking the third (depth) axis. This snapping ensures that subsequent drag operations map unambiguously to the two visible dimensions. Users can also explicitly toggle individual dimensions on or off: disabling one dimension collapses the 3D cube to a 2D plane; disabling two produces a 1D line. This scalable dimensionality lets users focus on specific trade-offs or examine the full multi-dimensional landscape as needed.

\subsubsection{Drag-Based Steering (Figure~\ref{fig:workflow}C)}
The central interaction paradigm for manipulating ideas in the evaluation space is \textit{drag-based steering}: users express how they want an idea to change by dragging its node to a new position in the evaluation space. When a user begins dragging, a ghost node connected by a dashed line visualizes the proposed trajectory from the original position to the target. This preview helps users understand exactly how the idea's scores will change before committing. Upon releasing the drag, the system offers two options: (1)~\textit{Drag-based iteration}---the system rewrites the idea's content to match the target position, producing a variant that embodies the new balance of trade-offs; or (2)~\textit{Evaluation correction}---the system updates the idea's scores without changing content, useful when users disagree with the AI's initial scoring. The steering interaction makes intent expression spatial and continuous rather than verbal and discrete, supporting rapid iteration.

\subsubsection{Idea Synthesis (Figure~\ref{fig:workflow}D)}
Beyond refining individual ideas, users often want to synthesize promising elements from multiple ideas. The system supports two combination mechanisms at different granularities:

\textbf{Drag-to-merge.} When a user drags one idea node close to another (within a proximity threshold), the target node highlights to indicate a potential merge. Releasing the drag triggers synthesis: the system combines the contributions of both parent ideas into a new hybrid idea that integrates their complementary strengths. The merged idea is then scored along all active dimensions and positioned accordingly in the evaluation space.

\textbf{Fragment incorporation.} Sometimes users identify a valuable phrase or concept within an idea that they want to incorporate elsewhere without adopting the entire idea. Users can extract text snippets from any idea and save them as fragments. To incorporate a fragment, users drag it onto any idea node; the system then revises the target idea by integrating the fragment's content. This supports non-linear ideation where users collect promising components throughout a session and recombine them flexibly.

\subsection{Implementation}

\subsubsection{Front-End: Multi-view Interaction and Shared State}
The front-end is implemented in React and maintains a shared idea graph where each node represents a candidate idea with (i) natural-language content, (ii) provenance links to parent ideas, and (iii) a score vector over the currently selected dimension pairs. The same shared state is rendered in two coordinated views: (1) Tree view. The tree provides provenance and supports growing the idea set by generating children, iterating an idea, or synthesizing two ideas. We render the tree as an interactive node-link diagram (D3).
(2) Dimensioned plot view. The plot visualizes the same nodes in a spatial evaluation space. For 3D interaction, we render the cube and nodes with a WebGL scene (Three.js via React Three Fiber), and provide camera rotation and face snapping to switch between overview and precise, plane-aligned drags.

To support richer workflows without overloading the evaluation space, we additionally implement fragment nodes: lightweight snippets created by selecting text in an idea card. Fragments are stored as separate nodes that can be dragged onto an idea to trigger a synthesis, allowing users to reuse partial intent without forcing every intermediate artifact to be fully evaluated.

\subsubsection{Back-End: LLM Primitives for Ideation and Evaluation}
The back-end is implemented as a Flask service that exposes a small set of LLM-powered primitives corresponding to the core UI operations:
\textit{Dimension generation} proposes candidate dimension pairs from a user intent.
\textit{Idea generation} produces seed ideas or child ideas.
\textit{Idea rewrite} produces an iterated idea conditioned on a user-specified directional shift in the evaluation space.
\textit{Idea synthesis} combines two ideas (or an idea and a fragment) into a unified proposal.
\textit{Evaluation} assigns each idea a vector of integer scores over up to three selected dimension pairs and provides brief reasoning for each score.

Prompt templates are maintained as editable YAML files so that interaction primitives and output schemas remain stable. In the evaluation prompt, each dimension pair is treated as a single spectrum with two opposing endpoints. We use a symmetric integer scale in [$-$50, 50], where $-$50 indicates full alignment with one pole of a trade-off dimension, +50 with the opposite, and 0 a balance, and map scores to 3D positions for visualization. When the user performs an explicit evaluation correction, the correction is recorded and can be incorporated into subsequent evaluation passes so that nearby comparisons remain coherent.

\subsection{Usage Scenario}
\label{sec:worked-example}

To illustrate how the four interactions compose into a coherent exploration session, we trace a usage scenario drawn from P02's study session (Figure~\ref{fig:workflow}). P02's research intent was ``using wearable data to train a multi-agent system for health prediction.''

\textbf{Dimension generation (Figure~\ref{fig:workflow}A).}
The system proposed five bipolar trade-off pairs. P02 selected three to define the evaluation space: \emph{Complex Models} vs.\ \emph{Simple Models} (X-axis), \emph{Data Privacy} vs.\ \emph{Data Utilization} (Y-axis), and \emph{Individual-Centric} vs.\ \emph{Population-Centric} (Z-axis). The remaining two pairs (Short-term Monitoring vs.\ Long-term Prediction, Single-Device vs.\ Multi-Device) were discarded. From this evaluative framework, the system generated three initial research ideas, each automatically scored along the three selected dimensions.

\textbf{Spatial navigation (Figure~\ref{fig:workflow}B).}
The three ideas were rendered as nodes in the 3D evaluation space. Node size encodes depth along the Z-axis: for example, \emph{Ethical Privacy Framework} (scored $-40$ on Individual--Population) appeared as a large, close node in the upper-left region (high Data Privacy, slightly Complex), while \emph{Real-Time Sensing Integration} (scored $0$ on Individual--Population) appeared smaller and further back in the lower-right (slight Data Utilization, slight Simple). P02 rotated the cube and snapped to the X--Y face to compare ideas on the Complex--Simple and Privacy--Utilization trade-offs.

\textbf{Drag-based steering (Figure~\ref{fig:workflow}C).}
Examining \emph{Agent-Based Modeling} (positioned far left on the Complex side), P02 wanted to explore a simpler, more data-driven variant. P02 dragged the node toward the Simple + Data Utilization quadrant. At the target position, the system generated a new idea, \emph{ML Real-Time Processing}, which retained the multi-agent framing but shifted toward lightweight, data-utilization-oriented algorithms.

\textbf{Idea synthesis (Figure~\ref{fig:workflow}D).}
P02 noticed that the newly generated \emph{ML Real-Time Processing} complemented the existing \emph{Real-Time Sensing Integration}, which focused on sensor data throughput. P02 dragged \emph{ML Real-Time Processing} onto \emph{Real-Time Sensing Integration} to merge them. The system produced \emph{Adaptive ML Real-Time Health}, a hybrid idea combining adaptive machine learning with real-time multi-source sensor fusion for wearable health monitoring.

%% file: sections/4-studyDesign.tex
\section{Study Design}
\label{sec:study-design}

We conducted a user study ($n=11$) to evaluate whether ResearchCube's spatial interactions support diverse ideation strategies and how researchers perceive the system's usability.

\subsection{Participants}
We recruited 11 participants (5F, 6M; age $M=25.5$, $SD=1.37$), all Master's or PhD students in CS/HCI/ECE-related fields with at least two peer-reviewed publications and at least two active research directions. Participants' research areas spanned human--robot interaction, accessibility, AI agents, language models, digital health, embodied AI, XR, and scholarly sensemaking (see Appendix~\ref{sec:demographics} for full demographics). Participants were recruited through email lists of HCI-related departments and research groups at multiple large research universities, academic social media, and snowball sampling. The study protocol was reviewed and approved by the Institutional Review Board (IRB) at the authors' affiliated university.

\subsection{Task}
Each participant performed an open-ended research ideation task using their own research topic. Starting from a self-provided research intent, participants used ResearchCube to (1)~select evaluative dimension pairs from AI-generated suggestions, (2)~generate and explore candidate ideas in the 3D evaluation space, (3)~iteratively refine ideas through drag-based steering, and (4)~synthesize ideas through synthesis and fragment incorporation.

\subsection{Procedure}
Sessions lasted approximately one hour and followed a structured protocol:

\begin{enumerate}
    \item Introduction (5~min): Participants provided informed consent, started recording and transcribing, then watched a tutorial video and an introduction document of ResearchCube's functions.
    \item Ideation session ($M=30.4$~min, $SD=13.4$): Participants used ResearchCube with their chosen research topic. They were encouraged to think aloud while exploring, evaluating, and refining ideas. Participants could end the session once they had tried all features and felt sufficiently familiar with the system. Actual durations varied by participant (range: 11--50~min). To ensure participant engagement, we asked them to select their three favorite ideas by the end of the session and explain their choices.
    \item Post-task assessment (5~min): Participants completed the System Usability Scale (SUS)~\cite{brookeSUSQuickDirty} questionnaire and ranked six system features by preference.
    \item Semi-structured interview (15~min):
    \begin{itemize}
        \item Interaction: Could you go through all the interactions, rank from best to worst, and comment?
        \item Presentation: What was your most intuitive or confusing moment when using the interface, and why?
        \item Real-world usage: How might you use this interface for your own research practice, and what additional features or improvements would make it indispensable?
    \end{itemize}
\end{enumerate}

Screen and audio recordings were captured throughout the session. Interviews were automatically transcribed and de-identified.

\subsection{Measures}
We collected three types of data:
\begin{itemize}
    \item Interaction logs: The system recorded all user actions, including drag events (start, end, and post-drag choice), idea generation, syntheses, fragment operations, 3D rotations, dimension toggles, and view changes.
    \item SUS questionnaire: We used the standard 10-item SUS~\cite{brookeSUSQuickDirty} for usability assessment, scored on the standard 0--100 scale.
    \item Screen and audio recordings: All sessions were screen- and audio-recorded to capture interaction behavior and think-aloud commentary.
    \item Interview transcripts: Semi-structured interviews were transcribed for qualitative analysis of user experience and ideation strategies.
\end{itemize}

Semi-structured interviews were analyzed using reflexive thematic analysis~\cite{braunUsingThematicAnalysis2006}. The first author served as primary analyst, conducting an initial open coding pass through all transcripts while reviewing screen recordings for additional behavioral context. Codes were then iteratively refined through discussion with the last author, resulting in the themes reported in Sections~\ref{sec:scaffolds}--\ref{sec:spatial-rep} (RQ1) and Sections~\ref{sec:drag}--\ref{sec:synthesis} (RQ2).

%% file: sections/5-results.tex
\section{Results}
\label{sec:results}

\subsection{Quantitative Overview}

\begin{table*}[t]
  \centering
  \small
  \caption{SUS item-level responses ($n=11$; scale: 1=Strongly Disagree, 5=Strongly Agree). Odd-numbered items are positively worded (higher is better); even-numbered items are negatively worded (lower is better).}
  \label{tab:sus-items}
  \renewcommand{\arraystretch}{1.15}
  \begin{tabular}{@{}clcc@{}}
    \toprule
    \textbf{\#} & \textbf{SUS Item} & \textbf{Mean} & \textbf{SD} \\
    \midrule
    Q1\textsuperscript{+} & I think that I would like to use this system frequently. & 3.82 & 0.98 \\
    Q2\textsuperscript{--} & I found the system unnecessarily complex. & 2.00 & 1.00 \\
    Q3\textsuperscript{+} & I thought the system was easy to use. & 3.91 & 0.94 \\
    Q4\textsuperscript{--} & I think that I would need the support of a technical person to be able to use this system. & 1.91 & 1.14 \\
    Q5\textsuperscript{+} & I found the various functions in this system were well integrated. & 4.45 & 0.69 \\
    Q6\textsuperscript{--} & I thought there was too much inconsistency in this system. & 1.36 & 0.67 \\
    Q7\textsuperscript{+} & I would imagine that most people would learn to use this system very quickly. & 3.73 & 1.19 \\
    Q8\textsuperscript{--} & I found the system very cumbersome to use. & 1.82 & 0.75 \\
    Q9\textsuperscript{+} & I felt very confident using the system. & 4.18 & 0.87 \\
    Q10\textsuperscript{--} & I needed to learn a lot of things before I could get going with this system. & 2.00 & 1.34 \\
    \midrule
    \multicolumn{2}{@{}l}{\textbf{Overall SUS Score}} & \textbf{77.5} & \textbf{15.9} \\
    \bottomrule
  \end{tabular}
\end{table*}

\begin{table}[t]
  \centering
  \small
  \caption{Feature rankings per participant (1=best, 6=worst; $n=11$). Tied ranks are mid-rank corrected (e.g., two features both ranked 1st receive rank 1.5). A Friedman test found no significant difference across features ($\chi^2(5)=1.54$, $p=.91$, Kendall's $W=.028$), reflecting high individual variability.}
  \label{tab:rankings}
  \begin{tabular}{lcccccc}
    \toprule
    & \textbf{Dim} & \textbf{Idea} & \textbf{3D} & \textbf{Drag} & \textbf{Synth.} & \textbf{Frag} \\
    \midrule
    P01 & 1.5 & 1.5 & 5 & 5 & 3 & 5 \\
    P02 & 3 & 6 & 1 & 2 & 5 & 4 \\
    P03 & 1 & 2 & 3.5 & 3.5 & 5 & 6 \\
    P04 & 3 & 3 & 6 & 3 & 3 & 3 \\
    P05 & 2 & 5 & 1 & 3 & 4 & 6 \\
    P06 & 5.5 & 5.5 & 3 & 1.5 & 4 & 1.5 \\
    P07 & 4 & 6 & 1 & 2 & 3 & 5 \\
    P08 & 1 & 2 & 6 & 3 & 5 & 4 \\
    P09 & 4 & 1 & 5 & 6 & 2 & 3 \\
    P10 & 5 & 1 & 6 & 4 & 2 & 3 \\
    P11 & 4 & 5 & 1 & 3 & 6 & 2 \\
    \midrule
    $\bar{x}$ & 3.1 & 3.5 & 3.5 & 3.3 & 3.8 & 3.9 \\
    \bottomrule
  \end{tabular}
\end{table}

\subsubsection{System Usability}

We computed SUS scores following the standard procedure~\cite{brookeSUSQuickDirty}:
\[
SUS = 2.5 \times \left(\sum_{i \in \text{odd}}(x_i - 1) + \sum_{i \in \text{even}}(5 - x_i)\right)
\]
where $x_i$ is the raw response (1--5) for item $i$. The overall SUS mean was 77.5 ($SD=15.9$, $Mdn=80.0$), exceeding the industry average of 68~\cite{bangorDeterminingWhatIndividual2009} and approaching the Grade~A threshold of 80.3~\cite{sauroQuantifyingUserExperience2016}. However, individual scores ranged from 45 (P08) to 100 (P07), a 55-point spread. Post-study interviews suggested that this polarization reflects differences in cognitive style: spatial thinkers rated the system highly, while text-oriented researchers found the 3D mapping cognitively burdensome.

Table~\ref{tab:sus-items} reports per-item SUS responses. The strongest positive signal was functional integration (Q5: $M=4.45$), indicating that participants perceived the system's components (dimension generation, spatial navigation, drag-based steering, and synthesis) as a coherent whole rather than disjoint features. Participants also reported high confidence (Q9: $M=4.18$) and low perceived inconsistency (Q6: $M=1.36$). The weakest positive item was learnability (Q7: $M=3.73$, $SD=1.19$), reflecting the highest variance among positively worded items; this aligns with the cognitive-style polarization noted above, where participants with weaker spatial reasoning required more onboarding. Similarly, the relatively elevated score on Q10 (needing to learn a lot, $M=2.00$, $SD=1.34$) showed the largest spread among negatively worded items, further suggesting that the learning curve varied substantially across participants.

\subsubsection{Feature Rankings}

Participants ranked six features: (a)~dimension generation, (b)~idea generation, (c)~spatial navigation, (d)~drag-based iteration, (e)~drag-to-merge, and (f)~fragment incorporation. Table~\ref{tab:rankings} summarizes individual rankings and mean ranks.

Dimension generation received the best mean rank ($\bar{x}=3.1$), with three participants (P01, P03, P08) placing it first. These participants valued how explicit trade-off dimensions scaffolded their evaluative thinking, a pattern explored further in Section~5.2.1. Drag-based iteration ranked second ($\bar{x}=3.3$), followed by idea generation and spatial navigation (tied at $\bar{x}=3.5$). Notably, Spatial navigation showed the highest polarization: it was ranked first by three participants (P02, P05, P07) who described themselves as spatial thinkers, but last by three others (P04, P08, P10) who found the 3D mapping cognitively burdensome, consistent with the SUS polarization above.

Idea synthesis ($\bar{x}=3.8$) and fragment incorporation ($\bar{x}=3.9$) were ranked lowest overall, with no participant ranking synthesis first. This quantitative result foreshadows the qualitative findings in Section~5.3.2, where participants identified issues with vagueness and exploration shrinkage after merging. Fragment incorporation was similarly polarized: P06 ranked it first (tied with drag), while P03 and P05 ranked it last.

A Friedman test found no significant difference in rankings across features ($\chi^2(5)=1.54$, $p=.91$, Kendall's $W=.028$). The very low Kendall's $W$ indicates near-zero concordance among participants, confirming that feature preferences were highly individualized, possibly shaped by individual cognitive styles, research domains, and ideation strategies rather than any universal ordering.

\subsection{RQ1: How Should a Research Ideation Space Be Designed So That Researchers Can Externalize, Compare, and Navigate Evaluative Trade-offs Among Candidate Ideas?}

Five themes emerged under RQ1 concerning the design of the ideation space.

\subsubsection{Dimensions as Externalization and Thinking Scaffolds}
\label{sec:scaffolds}
AI-generated trade-off dimensions served not merely as evaluation axes but as tools for externalizing and scaffolding researchers' thinking through three complementary mechanisms: 1) dimensions encouraged thinking when researchers felt stuck in ideation, 2) dimension generation prompted deeper reflection on evaluative criteria, and 3) spatial contextualization of dimensions helped offload working memory.

Dimensions provided an entry point for re-engaging with stalled ideation. P11 described wanting to use the system ``when I really can't figure something out, and I need something to give me a push.'' Explicit trade-off dimensions gave him a concrete starting point when internal reasoning had stalled.

The act of selecting and defining dimensions itself deepened evaluative reasoning. P01 described how dimension generation helped her think about her contributions in two ways:

\begin{quote}
``The dimension generation step mainly helped me in two ways. The first is that when it generates all these dimensions, these dimensions are essentially my contributions. So it can help me think about what vague, high-level categories of contribution there could be. And then when I go through them, I'd first eliminate some, like quantitative, statistical.

And then when I select the three dimensions I think are good, in order to compare better later, I definitely need to look at them carefully. So I'd go deeper into thinking. It's like my understanding of my own novelty has already reached 0.2. Then at this step, I probably want to push it to 0.8. So it would force me to think, under these levels, specifically which ones.'' (P01)
\end{quote}

P08 similarly observed that the suggested dimensions surfaced previously unconsidered perspectives: ``these three points (dimensions) in the parentheses are actually not bad, because I might not have thought too much about what dimensions I need to evaluate.''

Rendering ideas within a dimensionally structured space also helped offload working memory by making relational information visually accessible rather than requiring researchers to hold it in mind. P11 described how the 3D view reduced cognitive burden:

\begin{quote}
``It lets me better perceive each small idea's position within the entire [space] across all the dimensions. Rather than something that requires more mental effort. Human cognitive load is very easily overloaded. And a 3D environment can actually help distribute some of that. It offloads some working memory, transforming it into other types of things.'' (P11)
\end{quote}

\subsubsection{Fluid Dimensionality Spanning}
\label{sec:fluid}

The ideation space should fluidly span from single-dimension focus for targeted, non-tradeoff iteration, through two- and three-dimensional trade-off comparison, to even higher-dimensional exploration, capturing the full range of evaluative reasoning involved in research ideation. This fluidity is particularly important given participants' varying spatial thinking capabilities, which produced sharply divergent preferences for spatial navigation in the feature rankings (Table~\ref{tab:rankings}). As P02 noted, ``for me, I probably have better spatial imagination, so I like to directly operate in space. Some people might not.'' P01 similarly observed that ``I think both are still useful. In 3D, you can see an overall relationship, but in 1D it might be clearer.''

At the single-dimension end, focused iteration on one axis proved valuable for depth. P05 noted that ``I'd look at points where I want to specifically understand how something has developed, so my focus might only be on one side of a dimension,'' suggesting that researchers sometimes want to drill into a single pole rather than navigate the full trade-off spectrum.

At the multi-dimensional end, exposure to three simultaneous dimensions made participants want more rather than fewer. P11 reacted to the three-dimension limit with ``Can I only select three? That feels like such a pity.'' P10 proposed an engineering-inspired alternative: ``Actually what I think would be better is if you directly give a three-view diagram, one angle from this perspective, and then switch to this other perspective.'' Multiple participants (P01, P03, P09) affirmed the importance of more than two dimensions, as P09 put it, ``without 3D rotation, I can't see things clearly.''

\subsubsection{Dimensions as a Source of Agency and Control}
\label{sec:agency}
The spatial representation restored a sense of agency that participants found absent in conversational AI tools. P11 contrasted ResearchCube with chatbot-based ideation:

\begin{quote}
``I generally don't take my own research questions or what I currently have and directly ask AI, ask Claude or ask `what do you think about the direction I'm working on? What new ideas can you recommend? Or what papers can you recommend?' I have never done that. Because I feel those systems are having something that I completely cannot control making recommendations for me. But I think your current system gives me a feeling that I have some control. Over its idea generation process, I have control over it. I'm not letting AI [lead]. Rather, AI is just an aid to my own agency.'' (P11)
\end{quote}

However, a tension emerged between this sense of agency and the practical difficulty of defining dimensions. Users found AI-generated dimensions helpful as a starting point because creating dimensions from scratch proved cognitively demanding. P06 noted ``coming up with it on my own, sometimes it's also quite hard to describe what exactly I [want].'' P08 similarly struggled when trying to come up with a possible dimension: ``I can't really think of what its opposite would be.'' Even when participants sensed a dimension was off, modification remained difficult: ``I think this is a bit too off, but I also don't know how to modify it'' (P08). Yet users simultaneously desired more control: P01 needed to extensively edit inapplicable dimensions, P06 wanted to swap dimensions after exploration: ``after trying some things, I feel like there might be a dimension I'd rather look at. Because maybe when I initially chose the three, I might not have been sure. I had two that I was very sure were related to the idea, and then the remaining few I was actually unsure about, so I might just pick three first, and then try them out.'', and P08 wanted to assign different weights to dimensions: ``Is there a priority? Like, I want this one and this one.''

\subsubsection{Spatial Representation over Dashboard and Tree View}
\label{sec:spatial-rep}
Participants consistently preferred spatial encodings over both the numerical score dashboard and the tree view for evaluative comparison.

The score dashboard, which showed numeric scores for ideas on each evaluative dimension tradeoff (shown in Figure \ref{fig:teaser} C\&D), was the most frequently cited source of confusion, with six participants (P01, P02, P04, P06, P09, P11) reporting difficulties. Specific issues included: negative scores on bipolar dimensions were misread as poor evaluations rather than positions along a spectrum (P06); the swap button was misunderstood as swapping idea content rather than axis orientation (P04); and the dimensions panel was misinterpreted as a list of ideas rather than evaluation criteria (P09). P03 found the numerical scoring redundant: ``the scoring was a bit confusing. And I think there's honestly not much meaning to having a versus, because it repeats information.''

Beyond avoiding confusion, the spatial representation actively supported exploration. P10 described using the spatial layout to identify under-explored regions: ``clustered area, they all feel about the same, so I want to push it to other places and see if I can get better ideas. They're all actually in a similar flat plane. Basically sampling within this whole 3D space, seeing what results come out when you lean in different directions.'' These findings suggest that bipolar evaluations should be conveyed through spatial position and directional indicators rather than signed numbers.

The spatial representation was also preferred over the tree view for comparative evaluation. P11 explained:

\begin{quote}
``The results from the evaluation view are better than what I see directly from the tree view. This is much more intuitive. The biggest problem with this current [tree view] is that I can't see at a glance which [idea] is on which side [of the evaluation criteria]. But with this chart I clearly know, for example if I want to find things that are more [in one direction], I can go directly to the right side to see why this one is so special.'' (P11)
\end{quote}

P11 further noted that the two views serve complementary functions: ``this plot view is better suited for exploration,'' while the tree view is ``more useful when I'm done looking and come back to summarize.'' He observed that ``the tree view information is basically one-dimensional, I can only know the temporal, chronological order,'' whereas the plot view reveals ``evaluation criteria relationships'' that the tree view cannot show.

\subsection{RQ2: What Interaction Mechanisms Enable Researchers to Fluidly Iterate on and Synthesize Ideas Within Such a Space?}

We organize insights under RQ2 around two ideation interaction paradigms: drag-based steering and idea synthesis.

\subsubsection{Drag-Based Steering: Exploration, Cognitive Cost, and Input Constraints}
\label{sec:drag}
The construction of a dimensional space naturally invited participants to interact with it through drag-based steering. As P10 put it, drag-based steering felt like ``sampling within this whole 3D space, seeing what results come out when you lean in different directions.'' When ideas appeared ``clustered, they all feel about the same, so I want to push it to other places and see if I can get better ideas'' (P10). The spatial layout made under-explored regions visible, motivating participants to steer ideas into those gaps. P10 extended this logic further, asking ``Can I newly generate one in this area?'', suggesting that the canvas should support not only repositioning existing ideas but also generating new ones at specified positions in the space.

However, drag-based steering demanded nontrivial cognitive investment, particularly for participants less comfortable with spatial thinking. P08 observed: ``I still need to understand the 2D or 3D graph, and then drag it to where I want. That also takes some brainpower.'' Motivation was contingent on perceived information gain after each drag: ``every step, I want each one to have some benefits. If the direction I want to drag toward and the result it gives me don't match, then I probably wouldn't want to drag anymore'' (P08). This suggests that the system should provide immediate, visible information gain after each drag operation to sustain engagement.

A recurring limitation was the mismatch between 3D conceptual space and 2D input devices. P05 noted ``it can only detect that I'm dragging on a 2D plane, but the other dimension it doesn't know which direction I'm dragging,'' a constraint echoed by P10. This confirms the importance of disambiguation mechanisms such as face snapping or axis locking for making 3D manipulation tractable on conventional screens.

\subsubsection{Idea Synthesis and Fragment Incorporation}
\label{sec:synthesis}
Idea synthesis, including both full-idea merging and fragment-based recombination, received no strong preferences in the feature ranking (Table~\ref{tab:rankings}). Participants identified several challenges with the current merging workflow, including difficulty tracking what changed after a merge, loss of clarity when combining ideas, and narrowing of the exploration space through repeated convergence.

Since merging combines content from multiple ideas, the resulting text tends to be longer and denser, making it difficult to identify what is new. P08 reported: ``I feel there's so much text that it's hard to find what newly generated information it's presenting. When I look around, all these related keywords already appeared in the first round. And I want it to tell me something new, but that's hard to find, [I don't know] where to look. Everything seems to make sense.'' She proposed an audit trail to address this:

\begin{quote}
``Ideally, I want to see\ldots maybe my memory is really bad. So I want to know what my previous operation was. For example, you could put a note in front saying `I merged X and Y to generate this thing,' and then I could compare what information the previous one gave me, and what the newly generated thing is. That way I could directly grasp what's the new information generated by this tool.'' (P08)
\end{quote}

Beyond transparency, participants noted that merging could dilute previously crisp ideas: P08 observed ``I feel each individual thing is quite clear, but together they feel very vague.'' Repeated merging also narrowed exploration breadth: ``I think its quality is good at the beginning but it easily keeps converging. Your system keeps merging, converging things. And if I want something new, it can only be based on what we're currently discussing. Our stuff will go deeper and deeper step by step. In my project, there might also be a convergent-divergent cycle. I might want to switch to a different topic to try'' (P08). P02 observed that the value of merging was context-dependent: ``if you're cross-domain, cross-field merging it might be somewhat useful, but if you're on a relatively similar track, merging back and forth doesn't really make much sense.''

For the merging interaction itself, P05 suggested a magnetic snap feature to make the process more fluid: ``I don't know if you've considered making it a magnetic snap feature, like I drag this ball near the other one, and boop, it just snaps over by itself.''

%% file: sections/6-discussion_conclusion.tex
\section{Discussion}
\label{sec:discussion}

\subsection{Summary and Interpretation of Results}

Our study yielded several findings about how researchers interact with multi-dimensional trade-off spaces for ideation. Regarding RQ1 (how to design the research ideation space), bipolar trade-off dimensions proved to be more than evaluation axes. They functioned as cognitive scaffolds that externalized evaluative thinking and offloaded working memory (Section~\ref{sec:scaffolds}). Participants used them as prompts for deeper reflection on criteria they had not previously articulated, and the scaffolding effect was strongest when participants felt stuck. Rendering these dimensions spatially provided a sense of agency that participants contrasted with chatbot-based AI tools (Section~\ref{sec:agency}). When researchers can see and manipulate the evaluative structure that shapes AI-generated ideas, they feel in control of the generative process rather than passively receiving recommendations. Notably, dimension quality was perceived as higher than idea quality, likely because dimensions define a mapping space, a shared representation~\cite{heerAgencyAutomationDesigning2019} that researchers can reason about independently of any particular idea, whereas ideas are more context-dependent and require verification based on researchers' domain knowledge.

Participants also desired fluid transitions across dimensionality levels, from single-dimension focus to more than three dimensions simultaneously (Section~\ref{sec:fluid}). Rather than a fixed dimensional frame, the ideation space should be zoomable: narrowing to one axis for focused iteration, expanding to three for trade-off comparison, and potentially extending further through linked views or projections. A productive tension emerged between AI-suggested starting dimensions and users' evolving desire for control (Section~\ref{sec:agency}). Participants valued the cold-start scaffolding but increasingly wanted to edit, reorder, and regenerate dimensions as their understanding deepened, mirroring a broader pattern in human-AI collaboration where AI provides initial structure and the user incrementally formalizes and refines it.

For interaction mechanisms (RQ2), drag-based steering enabled participants to treat the dimensional space as a sampling environment (Section~\ref{sec:drag}), but demanded cognitive investment, particularly for participants less comfortable with spatial thinking. Idea synthesis faced challenges around content transparency, convergence, and context-dependence (Section~\ref{sec:synthesis}). Taken together, these findings validate that multi-dimensional research ideation tools, where trade-off dimensions are explicitly surfaced, spatially rendered, and directly manipulable, offer meaningful benefits over both unstructured chatbot interactions and fixed-scale evaluation interfaces.

\subsection{Limitations and Future Work}

\subsubsection{Speed and Quality Trade-off in Model Responses}
Direct intent manipulation relies on the feeling of immediacy: drag, release, and see the result. Our current implementation uses general-purpose LLMs, which introduce latency that can break this sense of direct manipulation. The broader point is that current models still lag behind the performance profile required for agency-preserving human-AI interactions like DIM~\cite{dingDirectIntentManipulation2025}: they need to be fast enough for fluid interaction while remaining faithful to the user's directional intent. Distilled or fine-tuned models optimized for instruction-following speed may help, though this trade-off may also resolve as foundation models improve.

\subsubsection{Spatial Cognition and Accessibility}
Spatial navigation produced the sharpest polarization in both SUS scores (45--100) and feature rankings (Table~\ref{tab:rankings}). This suggests that the cube may currently serve only participants with particular spatial cognition profiles. Future work should investigate adaptive interfaces that adjust dimensionality and representation based on user preferences, as well as non-spatial alternatives (e.g., tabular or narrative representations) for researchers who find 3D mapping cognitively burdensome.

\subsubsection{Transparent Idea Synthesis with Provenance}
Idea synthesis was the least preferred feature, and participants identified vagueness, convergence, and opacity as key issues. A notable limitation of the current implementation is that the merge prompt does not directly leverage dimensional information. It simply asks the LLM to ``combine the best aspects of both'' ideas, leaving the model to decide what is ``most valuable.'' Future work should incorporate dimensional context into synthesis (e.g., specifying which dimension to optimize in the merged result) and draw on prior work in analogical idea combination~\cite{srinivasanImprovingSelectionAnalogical2024} to give users more control over how ideas are composed. How to incorporate dimensionality into the merging process remains an open and interesting design question.

\subsubsection{Interactions Beyond Three Dimensions}
How to enable interaction in more than three spatial dimensions remains an open challenge. Several participants wanted more than three dimensions simultaneously, but the mapping from screen-based 2D input to higher-dimensional space is nontrivial. P10's proposal for three-view projections offers one starting point, as do approaches inspired by multi-layered geographic navigation (e.g., Google Earth). Future work should explore linked small-multiple views, focus-plus-context projections, more immersive space (e.g., AR/VR) and interactive dimension subsetting as mechanisms for navigating higher-dimensional trade-off spaces.

\subsection{Design Implications}

Beyond the specific system, our findings point to broader design implications for multi-dimensional research ideation tools and human-AI interfaces for complex knowledge work.

\subsubsection{Fluid Dimensionality and Dynamic Abstractions}
Our finding that participants wanted to fluidly switch between dimensional levels, and that no single ``best'' dimensionality exists, connects to a broader vision of dynamic abstractions as agency \cite{suhDynamicAbstractions2024}-preserving primitives for human-AI interaction. In ResearchCube, dimensions function as intermediate abstractions: they are neither raw data nor final outputs, but structured lenses through which researchers and AI jointly reason about trade-offs. The power of these abstractions lies in their dynamism: users can generate, select, edit, and discard them as understanding evolves.

This pattern resonates with recent work on dynamic abstractions in interactive systems~\cite{satyanarayanIntelligenceAsAgency2024}, where shared representations between human and AI are not fixed by the system designer but co-created through use. It also connects to the vision of malleable software, where the bottleneck is not the interface per se but the quality of the primitives users can define and manipulate~\cite{minMalleableOverviewDetailInterfaces2025}. The sense of power and control that participants reported from being able to define and manipulate their own evaluative dimensions echoes broader findings on artifact power in creative tools \cite{liBeyondArtifactPower2023}. Our results suggest that dimensions, as user-definable, AI-scaffolded abstractions, are a promising class of such primitives for research ideation.

\subsubsection{Progressive Dimensional Control and Incremental Formalization}
The tension between AI scaffolding and user control (Section~\ref{sec:agency}) implies that ideation systems should implement a progressive handoff model. Initially, the system proposes candidate dimensions to reduce cold-start burden. As the researcher's understanding evolves, the system should support editing individual dimensions, reordering or prioritizing them, regenerating dimensions while preserving user-curated ones, and adding user-defined dimensions alongside AI-generated ones.

This progression reflects a broader pattern of incremental formalization, the process by which initially informal, ambiguous representations become progressively more structured through iterative refinement. Participants' desire to revise dimensions mid-exploration (P06, P08) suggests that the formalization process should be ``loopy'' rather than linear: users move between exploring ideas and refining the evaluative framework itself. There may also be a hidden benefit to generating multiple dimensions at the outset (five pairs in our design, from which users select three): this leverages parallel prototyping~\cite{dowParallelPrototypingLeads2010}, giving users several candidate frameworks to compare before committing. This is especially valuable given the uncertainty about which dimensions are most relevant before exploration begins.

\section{Conclusion}

We presented ResearchCube, a multi-dimensional trade-off exploration system for AI-assisted research ideation that renders research ideas as manipulable nodes in a 3D evaluation space defined by user-selected bipolar dimensions. Through a qualitative study with 11 researchers, we identified findings organized around two research questions: under RQ1, we found that bipolar dimensions serve as cognitive scaffolds, participants desire fluid dimensionality spanning, spatial dimensions provide a sense of agency, and a productive tension exists between AI scaffolding and user control over dimensions; under RQ2, we found that drag-based steering enables spatial exploration but demands cognitive investment, and idea synthesis faces challenges around content transparency and convergence. Our findings suggest that spatializing evaluative trade-offs transforms AI-assisted ideation from passive consumption of AI recommendations into active, agency-preserving exploration. We discuss implications for the broader design of human-AI interfaces for complex knowledge work, including fluid dimensionality through dynamic abstractions and progressive dimensional control through incremental formalization.

%% file: sections/7-appendix.tex
\clearpage
\onecolumn
\appendix

\section{Participant Demographics}
\label{sec:demographics}

Table~\ref{tab:demographics} summarizes participant demographics and self-reported GenAI usage for research.

\begin{table}[h!]
  \centering
  \small
  \caption{Participant demographics and background ($n=11$). All participants confirmed at least two peer-reviewed publications.}
  \label{tab:demographics}
  \renewcommand{\arraystretch}{1.2}
  \begin{tabular}{@{}cccp{5cm}p{8cm}@{}}
    \toprule
    \textbf{PID} & \textbf{Age} & \textbf{Gender} & \textbf{Research Directions} & \textbf{GenAI for Research} \\
    \midrule
    P01 & 26 & F & Human--robot interaction; co-design in healthcare; AI tools for IDD care & Uses ChatGPT as a thinking partner to comment on rough ideas, outline directions, and surface keywords; does not rely on GenAI for idea generation or evaluation directly \\
    P02 & 24 & F & Video reasoning language models; wearable devices & Uses GenAI for literature review, idea generation, coding, and summarization \\
    P03 & 26 & F & Accessibility for blind users; accessibility for autistic children & Uses GenAI to proofread writing and organize fragmented ideas; does not use for research ideation \\
    P04 & 25 & F & Human behavior modeling with AI; proactive agents for healthcare; world modeling in finance & Uses ChatGPT and NotebookLM for coding, literature review, brainstorming, and grammar checks \\
    P05 & 24 & M & AI design for patients vs.\ AI design for healthcare professionals & Uses ChatGPT Edu for basic literature search, note-taking, and code debugging \\
    P06 & 25 & M & Automatic scientific discovery; social intelligent agents; multi-agent systems & Uses AI2 Asta for literature review; discusses papers with Gemini and ChatGPT for ideation; uses GenAI for coding and writing \\
    P07 & 25 & M & Large language models; language agents & Uses Gemini and GPT to chat about and evaluate research ideas; uses coding agents \\
    P08 & 26 & F & Digital health chatbots; multi-agent creativity and ideation; brain-inspired AI with HCI & Uses ChatGPT for ideation, literature collection, and research clarification; uses NotebookLM for resource gathering and informal learning \\
    P09 & 25 & M & ECE; RTL design; robotics & Uses GenAI for everything including coding and writing; does not use for idea generation \\
    P10 & 26 & M & Embodied AI; vision-language models & Uses GenAI for coding, debugging, technical report writing, and research idea exploration (ChatGPT thinking mode, Gemini) \\
    P11 & 29 & M & XR; scholarly sensemaking; hypertext & Uses GenAI for generating datasets, coding, and writing summaries; does not use for idea generation \\
    \bottomrule
  \end{tabular}
\end{table}

\section{System Prompts}
\label{sec:prompts}

This appendix documents the key prompts used in ResearchCube's backend. All prompts share a common system message:

\begin{tcolorbox}[breakable, colback=gray!5, colframe=gray!75!black, title=System Message, fonttitle=\bfseries]
\ttfamily\small
You are a distinguished senior professor, renowned for your visionary research and track record of field-shaping publications. Your goal is to conceive creative, high-impact research directions that can be realistically explored by your PhD students.
\end{tcolorbox}

\begin{tcolorbox}[breakable, colback=gray!5, colframe=gray!75!black, title=Diverse Idea Generation Prompt, fonttitle=\bfseries]
\ttfamily\small
Generate THREE creative and diverse research ideas based on the following intent. Each idea should take a fundamentally different approach to addressing the research problem.

\medskip
Intent: \{intent\}

\medskip
Additionally, based on recent literature, here are some related works that might inform your ideas: \{related\_works\_string\}

\medskip
For EACH of the three ideas, address: What is the problem? Provide a comprehensive description including background, current challenges, and why the issue persists. Include citations where relevant.

\medskip
DIVERSITY REQUIREMENT: The three ideas must be fundamentally different in their approach---use different methodologies, target different aspects, employ different techniques, and have different risk/reward profiles.

\medskip
Respond with THOUGHT (intuitions, motivations, how each idea differs) followed by a JSON array of three objects, each with ``Name'', ``Title'', and ``Problem'' fields.
\end{tcolorbox}

\begin{tcolorbox}[breakable, colback=gray!5, colframe=gray!75!black, title=Dimension Suggestion Prompt, fonttitle=\bfseries]
\ttfamily\small
Based on the following research intent, suggest 3 pairs of contrasting evaluation dimensions that would be most relevant for comparing research ideas.

\medskip
Research Intent: \{intent\}

\medskip
Each dimension pair should represent TWO OPPOSING ENDS OF THE SAME SPECTRUM. These are trade-offs, not independent dimensions.

\medskip
Examples: (HCI-oriented, AI-oriented); (Novel Methods, Established Frameworks); (Theoretical Depth, Practical Application); (High Risk High Reward, Low Risk Incremental).

\medskip
Requirements: (1) Each pair must represent opposing ends of a single spectrum. (2) Dimensions should be relevant to the research domain. (3) Dimensions should reflect meaningful trade-offs. (4) Provide clear descriptions for each end.

\medskip
Respond with THOUGHT followed by JSON containing ``dimension\_pairs'': an array of objects with ``dimensionA'', ``dimensionB'', ``descriptionA'', ``descriptionB'', and ``explanation'' fields.

\medskip
CRITICAL: A research idea can only be at one position on this spectrum (0--100), where 0 is completely dimensionA and 100 is completely dimensionB.
\end{tcolorbox}

\begin{tcolorbox}[breakable, colback=gray!5, colframe=gray!75!black, title=Idea Evaluation Prompt, fonttitle=\bfseries]
\ttfamily\small
Evaluate and score multiple research ideas for the following intent: \{intent\}

\medskip
Ideas to evaluate: \{ideas\}

\medskip
Dimension Pair 1: \{dimension\_pair\_1\_name\}

\quad-- \{dimension\_1\_a\} (Score: $-50$): \{dimension\_1\_a\_desc\}

\quad-- \{dimension\_1\_b\} (Score: $+50$): \{dimension\_1\_b\_desc\}

\medskip
Dimension Pair 2: \{dimension\_pair\_2\_name\}

\quad-- \{dimension\_2\_a\} (Score: $-50$): \{dimension\_2\_a\_desc\}

\quad-- \{dimension\_2\_b\} (Score: $+50$): \{dimension\_2\_b\_desc\}

\medskip
Requirements: For each idea, provide Dimension1Score ($-50$ to $+50$), Dimension2Score ($-50$ to $+50$), and brief reasoning for each score. Ensure ideas receive meaningfully different scores. Preserve exact original titles.
\end{tcolorbox}

\begin{tcolorbox}[breakable, colback=gray!5, colframe=gray!75!black, title=Idea Modification Prompt (Drag-Based Steering), fonttitle=\bfseries]
\ttfamily\small
Given a research idea and score adjustments based on user evaluation, generate a modified version that aligns with the specified target scores.

\medskip
Original Idea: \{idea\}

Score Adjustments: \{modifications\}

Research Intent: \{intent\}

\medskip
Guidelines: (1) Preserve the core research contribution unless extreme scores require fundamental transformation. (2) Adjust specific aspects (methodology, scope, resources) that would justify the target scores. (3) Ensure the modified idea remains aligned with the original intent. (4) Make concrete, actionable changes rather than superficial adjustments.

\medskip
Respond with THOUGHT (how you interpreted the adjustments) followed by MODIFIED IDEA JSON with the same structure as the original.
\end{tcolorbox}

\begin{tcolorbox}[breakable, colback=gray!5, colframe=gray!75!black, title=Idea Merging Prompt, fonttitle=\bfseries]
\ttfamily\small
Below are Idea A and Idea B, each produced by your doctoral team. Your task is to merge them into a new, feasible and more comprehensive idea.

\medskip
Idea A: \{idea\_a\}

Idea B: \{idea\_b\}

\medskip
Respond with THOUGHT (how you combined elements and why the merged result is stronger) followed by MERGED IDEA JSON with ``Name'', ``Title'', and ``Problem'' fields.
\end{tcolorbox}